\documentclass[aps,prb,twocolumn,eqsecnum,showpacs,superscriptaddress]{revtex4}
\usepackage{graphicx,color}
\usepackage{amsmath}
\usepackage{amssymb}
\usepackage{bm}

\newcommand{\Tr}{\mathop{\text{Tr}}\nolimits}
\newcommand{\ket}[1]{|{#1}\rangle}
\newcommand{\bra}[1]{\langle{#1}|}
\newcommand{\bras}[2]{{}_{#2}\hspace*{-0.2mm}\langle{#1}|}
\newcommand{\bracket}[2]{\langle#1|#2\rangle}
\newcommand{\ketbras}[3]{\ket{#1}_{#3}\hspace*{-0.2mm}\bra{#2}}

\begin{document}

\title{State Tomography of a Chain of Qubits Embedded in a Spin Field-Effect Transistor via Repeated Spin-Blockade Measurements on the Edge Qubit}



\author{Kazuya Yuasa}
\affiliation{Waseda Institute for Advanced Study, Waseda University, Tokyo 169-8050, Japan}

\author{Kosuke Okano}
\affiliation{Department of Physics, Waseda University, Tokyo 169-8555, Japan}

\author{Hiromichi Nakazato}
\affiliation{Department of Physics, Waseda University, Tokyo 169-8555, Japan}

\author{Saori Kashiwada}
\affiliation{Research Center for Integrated Quantum Electronics, Hokkaido University, Sapporo 060-8628, Japan}

\author{Kanji Yoh}
\affiliation{Research Center for Integrated Quantum Electronics, Hokkaido University, Sapporo 060-8628, Japan}



\date[]{March 19, 2009}

\begin{abstract}
As a possible physical realization of a quantum information processor, a system with stacked self-assembled InAs quantum dots buried in GaAs in adjacent to the channel of a spin field-effect transistor has been proposed.
In this system, only one of the stacked qubits, i.e.\ the \textit{edge} qubit (the qubit closest to the channel), is measurable via ``spin-blockade measurement.''
It is shown that the state tomography of the whole chain of the qubits is still possible even under such a restricted accessibility.
The idea is to make use of the entangling dynamics of the qubits.
A recipe for the two-qubit system is explicitly constructed and the effect of an imperfect fidelity of the measurement is clarified.
A general scheme for multiple qubits based on repeated measurements is also presented.
\end{abstract}


\maketitle

\section{Introduction}
\begin{figure}[b]
\includegraphics[width=0.47\textwidth]{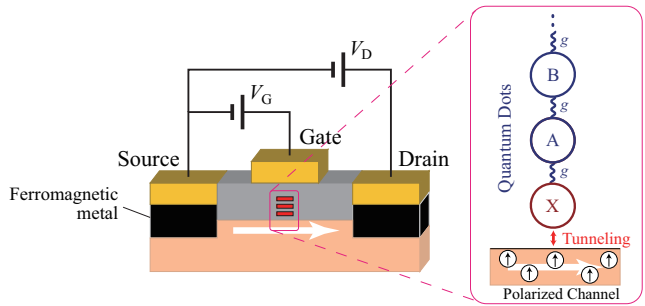}
\caption{(Color online) Spin FET embedded with quantum dots.}
\label{fig:FET}
\end{figure}

Towards realizations of quantum information processors,
a variety of physical systems have been proposed
and intensively investigated.
In particular, solid-state devices with the quantum bits (qubits) realized by the spins
of electrons confined in quantum dots in semiconductors \cite{ref:QdotTheories,ref:Qdots,ref:QdotsRMP}
are supposed to be promising in terms of its compatibility with existing semiconductor technology.
Among them, vertically stacked self-assembled InAs quantum dots have advantage of strong confinement of electrons which allows high temperature operation of the order of $1\,\text{K}$ as opposed to mK in confined 2DEG system by Schottky electrode.\cite{ref:Qdots,ref:QdotsRMP}
Thus, we have proposed and have been investigating, from both experimental
and theoretical aspects, a system with vertically stacked
self-assembled InAs dots buried in AlInAs barrier layer in adjacent to
the channel of a spin field-effect transistor (FET) (see Fig.\ \ref{fig:FET}).\cite{ref:qdot-PHYE,ref:KashiwadaYoh-JCG2007}

For quantum information processing, one should be able to perform initialization, quantum gate operations, and readout of the qubits.\cite{ref:DiVincenzoCriteria}
In the proposed setup depicted in Fig.\ \ref{fig:FET}, the qubits evolve under the interactions with the neighboring qubits and each of them would be rotated via electric spin resonance (ESR).
Furthermore, it is possible to measure repeatedly the state of the spin of the electron in the edge quantum dot, just above the channel of the FET, by making use of the ``spin-blockade effect,'' as will be recalled in Sec.\ \ref{sec:FET}\@.\cite{ref:qdot-PHYE,ref:KashiwadaYoh-JCG2007}

Although the other qubits than the one on the edge is not directly accessible by the proposed measurement scheme, one can still perform useful operations on the chain of qubits.
Multiple qubits can be initialized via repeated measurements only on the edge qubits,\cite{ref:qpfe,ref:qdot-PHYE} and an entanglement generation was discussed.\cite{ref:qdot-PHYE}
In the present article, we show that the \textit{state tomography} (or the \textit{state reconstruction}) is also available.

The determination of the quantum state is a highly nontrivial problem.\cite{ref:StateEstimationMISC}
A wave function, or more generally a density operator, of the state of a quantum system is not an observable and cannot be measured directly.
From a practical point of view, one can see the state only through measurable quantities.
In order to gain full information on the state of an $N$-level system, a list of the measured values (expectation values) of $N^2-1$ independent observables is required, with which all the matrix elements of its density matrix are \textit{reconstructed} and the \textit{tomography} of the state is accomplished.

The state tomography has been carried out for a variety of physical systems to analyze experiments.\cite{ref:TomoPhoton,ref:TomoWigner,ref:TomoAtom,ref:NMR,ref:TomoElecNucl2004,ref:TomoSuper,ref:TomoQdot,ref:TomoC60}
We are going to discuss the state tomography in the present setup.
One can measure only the edge qubit; still, it is possible to reconstruct the state of the \textit{whole} chain of the qubits.

\section{Spin FET Embedded with Quantum Dots}
\label{sec:FET}
The proposed device is illustrated in Fig.\ \ref{fig:FET}.\cite{ref:qdot-PHYE,ref:KashiwadaYoh-JCG2007}
A series of quantum dots is embedded in the FET structure, just above the channel.
A single electron is confined in each quantum dot and quantum information is encoded on its spin states, $\ket{\uparrow}$ and $\ket{\downarrow}$.
Such a situation where only a single electron is stored in each dot is realized by properly adjusting the gate voltage $V_\text{G}$.\cite{ref:QdotsRMP,ref:QdotsHoneycomb}
Each qubit would be rotated via ESR to perform single-qubit operations, and the qubits are entangled by the evolution under the interactions between the neighboring qubits.

The FET structure aims at measuring the spin state of the electron confined in the edge quantum dot X\@.
Notice first that one can detect the injection of an electron from the channel into the edge dot X by looking at the channel (source-drain) current $I_\text{D}$ as a function of the gate voltage $V_\text{G}$.
As the gate voltage $V_\text{G}$ is increased, the channel current $I_\text{D}$ increases. 
But if an electron tunnels from the channel into the edge dot in the meanwhile, the edge dot is charged and the channel current is suppressed.
As a result, the channel current $I_\text{D}$ drops down and exhibits a peak as a function of the gate voltage $V_\text{G}$ (Coulomb-blockade effect). 
Suppose now that the channel electrons are spin polarized in a definite spin state, say $\ket{\uparrow}$.
When the edge electron is in the state $\ket{\downarrow}$, a channel electron in $\ket{\uparrow}$ can tunnel into the edge dot at a certain gate voltage $V_\text{G}$, but, on the contrary, when the edge electron is in $\ket{\uparrow}$, the channel electron is not allowed to enter there due to Pauli's exclusion principle.
Therefore, if the polarized channel current $I_\text{D}$ drops down as the gate voltage $V_\text{G}$ is increased, one recognizes that the edge electron is in $\ket{\downarrow}_\text{X}$, while the growth of $I_\text{D}$ indicates that the edge electron is in $\ket{\uparrow}_\text{X}$. 
In this way, one can measure the spin state of the edge qubit X\@.
We call it ``spin-blockade measurement.''\cite{ref:qdot-PHYE,ref:KashiwadaYoh-JCG2007}

The feasibility of the present system is discussed in Ref.\ \onlinecite{ref:KashiwadaYoh-JCG2007}:
(i) the selective access to each individual qubit via ESR becomes possible by slightly modifying the compound ratio $x$ of the $\text{In}_{1-x}\text{Ga}_x\text{As}$ quantum dot, since the $g$ factor of the electron in a dot is altered in this way \cite{ref:gfactorArakawa} and the ESR absorption spectra of the electrons in different dots can be separated;
(ii) the strength of the exchange interaction energy between qubits and the corresponding characteristic time scale are estimated, showing the feasibility;
(iii) the modulation of the channel current by single electron charging in a quantum dot adjacent to the channel has been demonstrated in a trial structure with a single layer of quantum dot, with unpolarized channel current.
Spin-polarized channel will be available by replacing the normal metal electrode with a ferromagnet, candidate materials for which are being intensively investigated: see Ref.\ \onlinecite{ref:YohFerhat}.
Spin decoherence time has been reported to be $2\,\text{ns}$ for InAs self-assembled quantum dots,\cite{ref:DecoInAs} while the spin-flipping time is estimated to be $50\,\text{ps}$.\cite{ref:KashiwadaYoh-JCG2007}

One of the remarkable features of the spin-blockade measurement is that one can measure $\ket{\uparrow}_\text{X}$ \textit{repeatedly}.
Although one is allowed to measure only the edge qubit X, this feature enables one to perform useful operations on the chain of qubits.
The initialization of multiple qubits via repeated measurements on the edge qubit and an entanglement generation are discussed in Refs.\ \onlinecite{ref:qdot-PHYE} and \onlinecite{ref:qpfe}.
Furthermore, the present article clarifies that the \textit{tomography} of the state of the \textit{whole} qubits X+A+B+$\cdots$ is also possible via the spin-blockade measurements only on the \textit{edge} qubit X.

\section{The Idea}
\label{sec:idea}
In the proposed setup, it is possible to measure the state $\ket{\uparrow}_\text{X}$ of the edge qubit X\@.
But at the same time, the measurement of $\ket{\downarrow}_\text{X}$ implies that a channel electron in $\ket{\uparrow}$ has been injected into the edge dot and the edge qubit X has been destroyed.
Furthermore, other qubits than X are out of the reach of the spin-blockade measurement.
Still, there is a way to get the full information about the state of the whole chain of the qubits.

To measure different states of X from $\ket{\uparrow}_\text{X}$ without loosing it by the injection of a channel electron, we apply a spin rotation just before a spin-blockade measurement.
For instance, finding X in $\ket{\uparrow}_\text{X}$ just after rotating X by the angle $\pi/2$ around the $y$ axis is essentially the measurement of the spin X oriented in the $x$ direction [a superposed state $(\ket{\uparrow}_\text{X}+\ket{\downarrow}_\text{X})/\sqrt{2}$] before the rotation.
In this way, it is possible to measure any state of X\@.

The idea for getting information about the states of other qubits than X is to make use of the entangling dynamics of the chain of the qubits and the collapse of the state by the measurement on X\@.
Although we are allowed to measure only X, such measurement would reflect the state of the other qubits due to the entanglement between X and the rest\@.

The strategy for the state tomography of the whole chain of the qubits is therefore the following.
The qubits X+A+B+$\cdots$ evolve, from a given state $\varrho$ to be reconstructed, under the action of the Hamiltonian
\begin{equation}
H=g_\text{XA}\bm{\sigma}^\text{(X)}\cdot\bm{\sigma}^\text{(A)}
+g_\text{AB}\bm{\sigma}^\text{(A)}\cdot\bm{\sigma}^\text{(B)}
+\cdots,
\label{eqn:Hamiltonian}
\end{equation}
where $\bm{\sigma}^\text{(Q)}$ represents the spin operator of qubit $\text{Q}\,(=\text{X},\text{A},\text{B},\ldots)$.
During the evolution, we rotate some qubits and measure $\ket{\uparrow}_\text{X}$ a few times at definite timings according to a certain recipe.
We prepare the same initial state $\varrho$ and perform such a fixed sequence of operations many times to obtain the probability for every measurement in the sequence to find X in the state $\ket{\uparrow}_\text{X}$.
(Notice that the measurement at the end of a sequence can be $\ket{\downarrow}_\text{X}$, since we do not need to proceed further.)

Consider, for instance, the following series of operations: (rotation of X by an angle $\theta$ around the $x$ axis) $\to$ (wait for time $\tau$) $\to$ (measurement of $\ket{\uparrow}_\text{X}$) $\to$ $\cdots$ $\to$ (measurement of $\ket{\downarrow}_\text{X}$).
The probability of getting the relevant result at every measurement implemented in the series is given by
\begin{align}
p
&=\Tr\{
P_\downarrow\cdots P_\uparrow U(\tau)R_x^\text{(X)}(\theta)\varrho 
R_x^{\text{(X)\dag}}(\theta)U^\dag(\tau)P_\uparrow\cdots P_\downarrow
\}
\nonumber\displaybreak[0]\\
&=\Tr\{
\varrho 
[R_x^{\text{(X)\dag}}(\theta)U^\dag(\tau)P_\uparrow\cdots P_\downarrow
\cdots P_\uparrow U(\tau)R_x^\text{(X)}(\theta)]
\},
\label{eqn:prob}
\end{align}
where $R_i^\text{(Q)}(\theta)=e^{-\frac{i}{2}\theta\sigma_i^\text{(Q)}}$ is the operator that rotates qubit $\text{Q}\,(=\text{X},\text{A},\text{B},\ldots)$ by an angle $\theta$ around the $i\,(=x,y)$ axis, $P_{{\uparrow}({\downarrow})}=\ketbras{{\uparrow}({\downarrow})}{{\uparrow}({\downarrow})}{\text{X}}$ is the projection operator that represents the measurement of $\ket{{\uparrow}({\downarrow})}_\text{X}$, and $U(\tau)=e^{-iH\tau}$ is the time-evolution operator between two operations and entangles the chain of the qubits.

This is the quantity that we can measure in the present system.
It can be viewed as the expectation value $p=\Tr\{\varrho\mathcal{O}\}$ of a Hermitian operator $\mathcal{O}$ in the relevant state $\varrho$, where $\mathcal{O}$ consists of $P_\uparrow$, $U(\tau)$, $R_x^{(\text{X})}(\theta)$, and so on.
Therefore, by suitably arranging the sequences of the operations, we can collect the sufficient number ($4^M-1$ for an $M$-qubit system) of expectation values of linearly independent operators that allow us to reconstruct the given state $\varrho$.

\begin{table}
\caption{Fifteen linearly independent sequences of operations sufficient to reconstruct a state of two qubits.
$g=g_\text{XA}$ is the coupling constant between the two qubits.}
\label{tab:SeqTwoQubits}
\begin{tabular}{ll}
\hline
\hline
Probs.&Sequences of Operations\\\hline
\smallskip
$p^{(1)}_{{\uparrow}({\downarrow})}$
&
$
P_\uparrow
\to U(\frac{\pi}{4g})
\to P_{{\uparrow}({\downarrow})}
$\\
\smallskip
$p^{(2)}_\downarrow$
& 
$
U(\frac{\pi}{4g})
\to P_\uparrow 
\to U(\frac{\pi}{4g})
\to P_\downarrow 
$\\
\smallskip
$p^{(3)}_\uparrow$
&
$
R_y(\pi)
\to P_\uparrow
\to U(\frac{\pi}{4g})
\to P_\uparrow  
$\\
\smallskip
$p^{(4)}_{{\uparrow}({\downarrow})}$
& 
$
P_\uparrow
\to U(\frac{\pi}{4g})
\to R_y(\frac{\pi}{2})
\to P_{{\uparrow}({\downarrow})}
$\\
\smallskip
$p^{(5)}_{{\uparrow}({\downarrow})}$
&
$
P_\uparrow
\to U(\frac{\pi}{4g})
\to R_x(\frac{\pi}{2})
\to P_{{\uparrow}({\downarrow})}
$\\
\smallskip
$p^{(6)}_{{\uparrow}({\downarrow})}$
&
$
U(\frac{\pi}{4g})
\to P_\uparrow 
\to U(\frac{\pi}{4g})
\to R_y(\frac{\pi}{2})
\to P_{{\uparrow}({\downarrow})}
$\\
\smallskip
$p^{(7)}_{{\uparrow}({\downarrow})}$
&
$
U(\frac{\pi}{4g})
\to P_\uparrow 
\to U(\frac{\pi}{4g})
\to R_x(\frac{\pi}{2})
\to P_{{\uparrow}({\downarrow})}
$\\
\smallskip
$p^{(8)}_{{\uparrow}({\downarrow})}$
&
$
R_y(\pi)
\to U(\frac{\pi}{4g})
\to P_\uparrow 
\to U(\frac{\pi}{4g})
\to R_y(\frac{\pi}{2})
\to P_{{\uparrow}({\downarrow})}
$\\
\smallskip
$p^{(9)}_{{\uparrow}({\downarrow})}$
&
$
R_y(\pi)
\to U(\frac{\pi}{4g})
\to P_\uparrow 
\to U(\frac{\pi}{4g})
\to R_x(\frac{\pi}{2})
\to P_{{\uparrow}({\downarrow})}
$\\
\smallskip
$p^{(10)}_{{\uparrow}({\downarrow})}$
&
$
R_y(\pi)
\to P_\uparrow 
\to U(\frac{\pi}{4g})
\to R_y(\frac{\pi}{2})
\to P_{{\uparrow}({\downarrow})}
$\\
\smallskip
$p^{(11)}_{{\uparrow}({\downarrow})}$
&
$
R_y(\pi)
\to P_\uparrow 
\to U(\frac{\pi}{4g})
\to R_x(\frac{\pi}{2})
\to P_{{\uparrow}({\downarrow})}
$\\
\smallskip
$p^{(12)}_{{\uparrow}({\downarrow})}$
&
$
R_y(\frac{\pi}{2})
\to P_\uparrow 
\to U(\frac{\pi}{4g})
\to P_{{\uparrow}({\downarrow})}
$\\
\smallskip
$p^{(13)}_{{\uparrow}({\downarrow})}$
&
$
R_x(\frac{\pi}{2})
\to P_\uparrow 
\to U(\frac{\pi}{4g})
\to P_{{\uparrow}({\downarrow})}
$\\
\smallskip
$p^{(14)}_{{\uparrow}({\downarrow})}$
&
$
R_y(\frac{\pi}{2})
\to P_\uparrow 
\to U(\frac{\pi}{4g})
\to R_x(\frac{\pi}{2})
\to P_{{\uparrow}({\downarrow})}
$\\
$p^{(15)}_{{\uparrow}({\downarrow})}$
&
$
R_x(\frac{\pi}{2})
\to P_\uparrow 
\to U(\frac{\pi}{4g})
\to R_y(\frac{\pi}{2})
\to P_{{\uparrow}({\downarrow})}
$
\\\hline\hline
\end{tabular}
\end{table}
\begin{table}
\caption{Reconstruction of the two-qubit state $\varrho$ from the probabilities $p^{(n)}_{{\uparrow}({\downarrow})}$ obtained by the sequences of operations listed in Table \ref{tab:SeqTwoQubits}\@.
Here, $\ket{\uparrow\downarrow}=\ket{\uparrow}_\text{X}\ket{\downarrow}_\text{A}$, etc.}
\label{tab:RhoTwoQubits}
\begin{tabular}{r@{}r@{}l}
\hline
\hline
\smallskip
$\bra{\uparrow\uparrow}\varrho\ket{\uparrow\uparrow}={}$&
\multicolumn{1}{l}{$p^{(1)}_\uparrow$,}&
\qquad
$\bra{\downarrow\downarrow}\varrho\ket{\downarrow\downarrow}
=p^{(3)}_\uparrow$
\\
\smallskip
$\bra{\uparrow\downarrow}\varrho\ket{\uparrow\downarrow}={}$&
\multicolumn{1}{l}{$p^{(1)}_\downarrow$,}&
\qquad
$\bra{\downarrow\uparrow}\varrho\ket{\downarrow\uparrow}
=p^{(2)}_\downarrow$
\\
$\bra{\uparrow\uparrow}\varrho\ket{\uparrow\downarrow}={}$&
$-\frac{1}{2}\,\Bigl($&
$p^{(4)}_\uparrow-p^{(4)}_\downarrow\Bigr)
-\frac{i}{2}\,\Bigl(p^{(5)}_\uparrow-p^{(5)}_\downarrow\Bigr)
$\\
$\bra{\uparrow\downarrow}\varrho\ket{\uparrow\uparrow}={}$&
$-\frac{1}{2}\,\Bigl($&
$p^{(4)}_\uparrow-p^{(4)}_\downarrow\Bigr)
+\frac{i}{2}\,\Bigl(p^{(5)}_\uparrow-p^{(5)}_\downarrow\Bigr)
$\\
$\bra{\uparrow\uparrow}\varrho\ket{\downarrow\uparrow}={}$&
$-\frac{1}{2}\,\Bigl($&
$p^{(6)}_\uparrow-p^{(6)}_\downarrow\Bigr)
-\frac{i}{2}\,\Bigl(p^{(7)}_\uparrow-p^{(7)}_\downarrow\Bigr)
$\\
$\bra{\downarrow\uparrow}\varrho\ket{\uparrow\uparrow}={}$&
$-\frac{1}{2}\,\Bigl($&
$p^{(6)}_\uparrow-p^{(6)}_\downarrow\Bigr)
+\frac{i}{2}\,\Bigl(p^{(7)}_\uparrow-p^{(7)}_\downarrow\Bigr)
$\\
$\bra{\downarrow\downarrow}\varrho\ket{\uparrow\downarrow}={}$&
$\frac{1}{2}\,\Bigl($&
$p^{(8)}_\uparrow-p^{(8)}_\downarrow\Bigr)
+\frac{i}{2}\,\Bigl(p^{(9)}_\uparrow-p^{(9)}_\downarrow\Bigr)
$\\
$\bra{\uparrow\downarrow}\varrho\ket{\downarrow\downarrow}={}$&
$\frac{1}{2}\,\Bigl($&
$p^{(8)}_\uparrow-p^{(8)}_\downarrow\Bigr)
-\frac{i}{2}\,\Bigl(p^{(9)}_\uparrow-p^{(9)}_\downarrow\Bigr)
$\\
$\bra{\downarrow\downarrow}\varrho\ket{\downarrow\uparrow}={}$&
$\frac{1}{2}\,\Bigl($&
$p^{(10)}_\uparrow-p^{(10)}_\downarrow\Bigr)
+\frac{i}{2}\,\Bigl(p^{(11)}_\uparrow-p^{(11)}_\downarrow\Bigr)
$\\
$\bra{\downarrow\uparrow}\varrho\ket{\downarrow\downarrow}={}$&
$\frac{1}{2}\,\Bigl($&
$p^{(10)}_\uparrow-p^{(10)}_\downarrow\Bigr)
-\frac{i}{2}\,\Bigl(p^{(11)}_\uparrow-p^{(11)}_\downarrow\Bigr)
$\\
$\bra{\uparrow\uparrow}\varrho\ket{\downarrow\downarrow}={}$&
$\frac{1}{4}\,\Bigl($&
$\langle\sigma_x^\text{(X)}\sigma_x^\text{(A)}\rangle
-\langle\sigma_y^\text{(X)}\sigma_y^\text{(A)}\rangle
$\\
&&\multicolumn{1}{r}{${}-i\langle\sigma_x^\text{(X)}\sigma_y^\text{(A)}\rangle
-i\langle\sigma_y^\text{(X)}\sigma_x^\text{(A)}\rangle
\Bigr)
$}\\
$\bra{\downarrow\downarrow}\varrho\ket{\uparrow\uparrow}={}$&
$\frac{1}{4}\,\Bigl($&
$\langle\sigma_x^\text{(X)}\sigma_x^\text{(A)}\rangle
-\langle\sigma_y^\text{(X)}\sigma_y^\text{(A)}\rangle
$\\
&&\multicolumn{1}{r}{${}+i\langle\sigma_x^\text{(X)}\sigma_y^\text{(A)}\rangle
+i\langle\sigma_y^\text{(X)}\sigma_x^\text{(A)}\rangle
\Bigr)
$}\\
$\bra{\uparrow\downarrow}\varrho\ket{\downarrow\uparrow}={}$&
$\frac{1}{4}\,\Bigl($&
$\langle\sigma_x^\text{(X)}\sigma_x^\text{(A)}\rangle
+\langle\sigma_y^\text{(X)}\sigma_y^\text{(A)}\rangle
$\\
&&\multicolumn{1}{r}{${}+i\langle\sigma_x^\text{(X)}\sigma_y^\text{(A)}\rangle
-i\langle\sigma_y^\text{(X)}\sigma_x^\text{(A)}\rangle
\Bigr)
$}\\
$\bra{\downarrow\uparrow}\varrho\ket{\uparrow\downarrow}={}$&
$\frac{1}{4}\,\Bigl($&
$\langle\sigma_x^\text{(X)}\sigma_x^\text{(A)}\rangle
+\langle\sigma_y^\text{(X)}\sigma_y^\text{(A)}\rangle
$\\
\smallskip
&&\multicolumn{1}{r}{${}-i\langle\sigma_x^\text{(X)}\sigma_y^\text{(A)}\rangle
+i\langle\sigma_y^\text{(X)}\sigma_x^\text{(A)}\rangle
\Bigr)
$}
\\
$\langle\sigma_x^\text{(X)}\sigma_x^\text{(A)}\rangle={}$&
$2\,\Bigl($&
$p^{(12)}_\uparrow-p^{(12)}_\downarrow
\Bigr)+\langle\sigma_x^\text{(A)}\rangle
$\\
$\langle\sigma_y^\text{(X)}\sigma_y^\text{(A)}\rangle={}$&
$2\,\Bigl($&
$p^{(13)}_\uparrow-p^{(13)}_\downarrow
\Bigr)-\langle\sigma_y^\text{(A)}\rangle
$\\
$\langle\sigma_x^\text{(X)}\sigma_y^\text{(A)}\rangle={}$&
$-2\,\Bigl($&
$p^{(14)}_\uparrow-p^{(14)}_\downarrow
\Bigr)+\langle\sigma_y^\text{(A)}\rangle
$\\
\smallskip
$\langle\sigma_y^\text{(X)}\sigma_x^\text{(A)}\rangle={}$&
$-2\,\Bigl($&
$p^{(15)}_\uparrow-p^{(15)}_\downarrow
\Bigr)-\langle\sigma_x^\text{(A)}\rangle
$
\\
$\langle\sigma_x^\text{(A)}\rangle={}$&
$-\Bigl($&
$p^{(4)}_\uparrow-p^{(4)}_\downarrow\Bigr)
+\Bigl(p^{(10)}_\uparrow-p^{(10)}_\downarrow\Bigr)
$\\
$\langle\sigma_y^\text{(A)}\rangle={}$&
$\Bigl($&
$p^{(5)}_\uparrow-p^{(5)}_\downarrow\Bigr)
-\Bigl(p^{(11)}_\uparrow-p^{(11)}_\downarrow\Bigr)
$\hspace*{17.9mm}
\\
\hline
\hline
\end{tabular}
\end{table}
Let us demonstrate how the above procedure works in the simplest case, for two qubits X+A\@.
Fifteen linearly independent sequences of operations that are sufficient to reconstruct a two-qubit state $\varrho$ are listed in Table \ref{tab:SeqTwoQubits}, where $R_i(\theta)=R_i^\text{(X)}(\theta)R_i^\text{(A)}(\theta)$ is a ``global'' rotation, which rotates both X and A at the same time (we can also construct a recipe with ``local'' rotations, which act separately on X or A).
Note that $p_\uparrow^{(n)}$ and $p_\downarrow^{(n)}$ are obtained simultaneously, by a common ensemble of experimental data collected for the $n$th sequence of operations.
By substituting the probabilities $p_{{\uparrow}({\downarrow})}^{(n)}$ measured for the sequences of operations into the formulas listed in Table \ref{tab:RhoTwoQubits}, all of the $16$ matrix elements of the two-qubit state $\varrho$ are disclosed and the state $\varrho$ is reconstructed.

\begin{figure}[t]
\begin{tabular}{ll}
(a) for $p_\uparrow^{(1)}$&(b) for $p_\downarrow^{(1)}$\\
\includegraphics[width=0.23\textwidth]{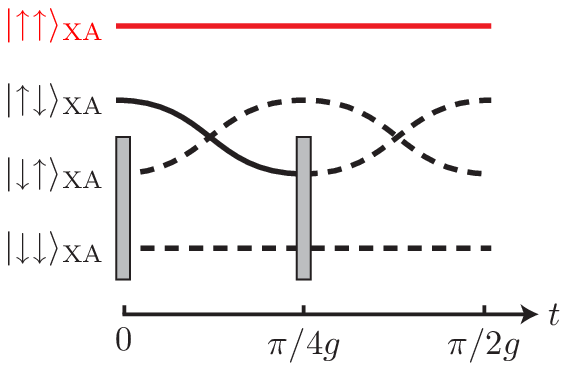}&
\includegraphics[width=0.23\textwidth]{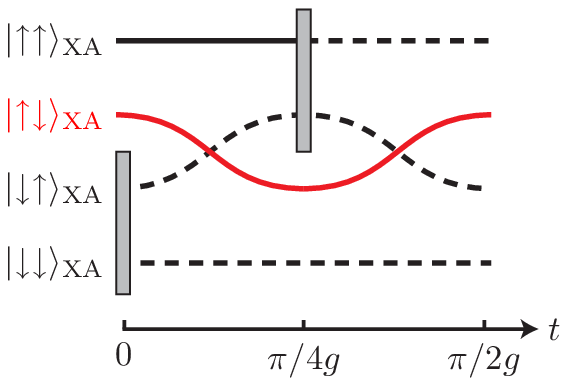}
\end{tabular}
\begin{tabular}{l}
(c) for $p_\downarrow^{(2)}$\\
\includegraphics[width=0.23\textwidth]{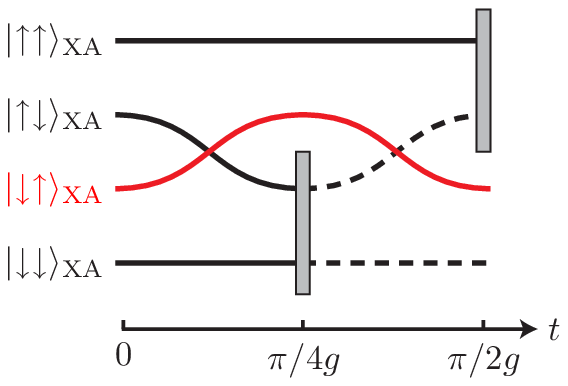}
\end{tabular}
\caption{(Color online) Mechanism of the tomography, for the sequences of operations for (a) $p_\uparrow^{(1)}$, (b) $p_\downarrow^{(1)}$, and (c) $p_\downarrow^{(2)}$.
The curves describe how the components of a state evolve and entangle in time according to the Hamiltonian (\ref{eqn:Hamiltonian}), while the ``walls'' represent the projective measurements on qubit X\@.
The survived curves which are not shut off by the walls correspond to the events where every measurement gives the desired result.
The probabilities of such events are nothing but the probabilities $p_{{\uparrow}({\downarrow})}^{(n)}$ for the sequences of operations and are equivalent (in the simple cases shown here) to the occupations of the survived states in the given initial state $\varrho$, yielding its relevant matrix elements, (a) $\bra{\uparrow\uparrow}\varrho\ket{\uparrow\uparrow}$, (b) $\bra{\uparrow\downarrow}\varrho\ket{\uparrow\downarrow}$, and (c) $\bra{\downarrow\uparrow}\varrho\ket{\downarrow\uparrow}$.}
\label{fig:TomoMech}
\end{figure}

The mechanism of the tomography is understood as follows, for relatively simple cases.
The curves in Fig.\ \ref{fig:TomoMech} describe the entangling dynamics according to the Hamiltonian (\ref{eqn:Hamiltonian}).
If the result of a measurement on X is ``no,'' two of the four components of the state of X+A are projected out.
The survival through such projections corresponds to the event where every measurement in a sequence gives ``yes,'' and the probability of such an event rephrases how much the survived component was contained in the given initial state $\varrho$.
The sequences are designed so as to reflect all of the matrix elements of $\varrho$, and the inversion of a linear relationship between the probabilities and the matrix elements reconstructs the given state $\varrho$.

\section{Partially Polarized Spin Channel}
\label{sec:PartialPolarization}
The realization of the polarized spin channel is one of the important issues to be tackled.\cite{ref:YohFerhat}
If the channel is only partially polarized, the fidelity of the spin-blockade measurement on X is degraded and the performance of the tomography deteriorates.
Let us clarify the effect of the partially polarized spin channel on the tomographic scheme presented in the previous section.

When the spin polarization is not perfect but $r$, the state of the electron in the channel would be effectively described by the density operator
\begin{align}
\varrho_\text{ch}
&=r\ket{\uparrow}\bra{\uparrow}+\frac{1-r}{2}\openone
\nonumber\displaybreak[0]\\
&=\frac{1+r}{2}\ket{\uparrow}\bra{\uparrow}
+\frac{1-r}{2}\ket{\downarrow}\bra{\downarrow}.
\label{eqn:PartialPolarization}
\end{align}
Due to the presence of the undesired ingredient, $\ket{\downarrow}\bra{\downarrow}$ in (\ref{eqn:PartialPolarization}), the confirmation of the increase in the channel current in response to the sweep of the gate voltage does not result in the pure projection $\varrho\to P_\uparrow\varrho P_\uparrow$ but would induce 
\begin{equation}
\varrho
\to\frac{1+r}{2}P_\uparrow\varrho P_\uparrow
+\frac{1-r}{2}P_\downarrow\varrho P_\downarrow,
\label{eqn:NonidealMeas}
\end{equation}
reflecting the classical mixture of the two states in (\ref{eqn:PartialPolarization}) (with other possible non-ideal features omitted).\cite{note:Projection}

\begin{figure}
\begin{tabular}{cc}
\includegraphics[width=0.233\textwidth]{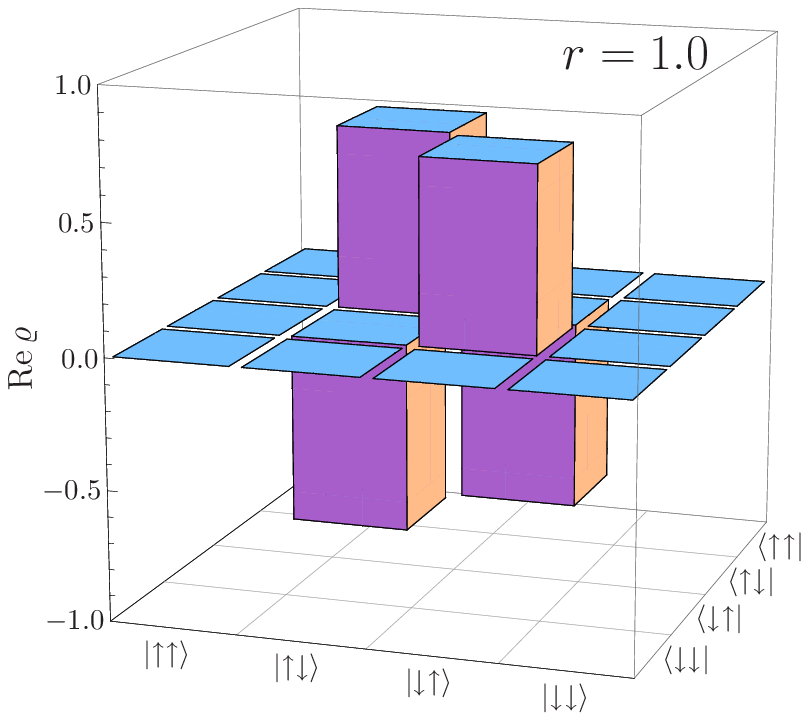}&
\includegraphics[width=0.233\textwidth]{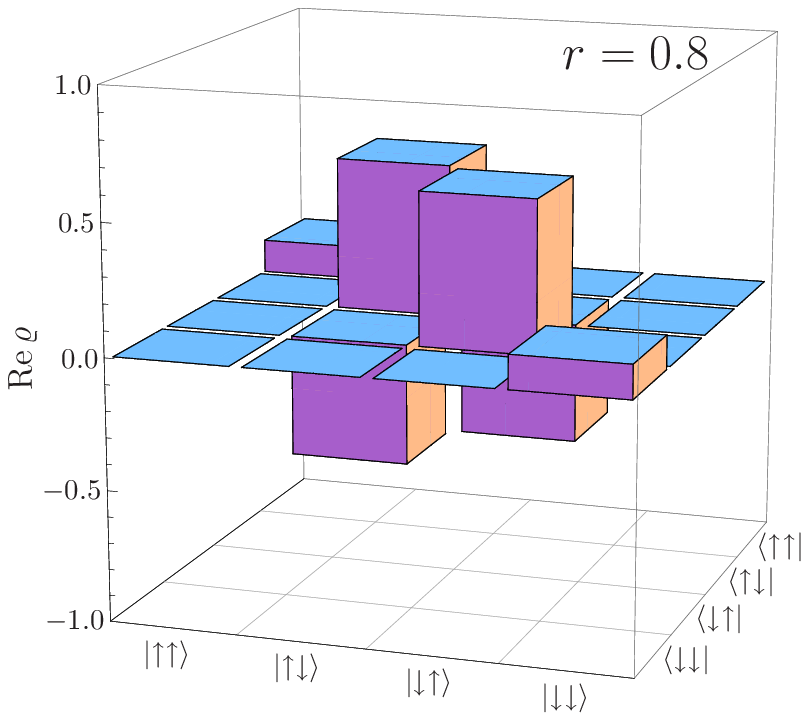}\\
\includegraphics[width=0.233\textwidth]{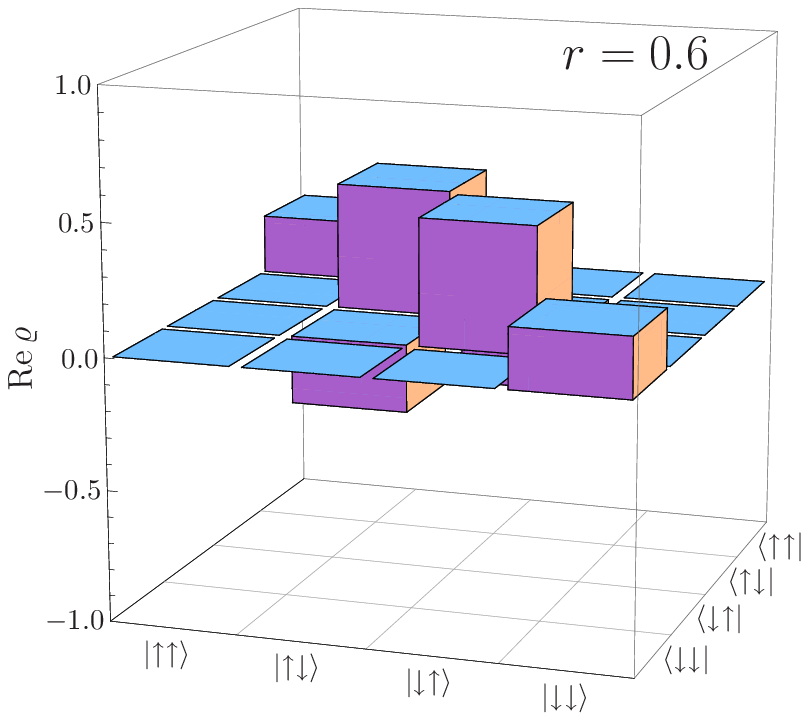}&
\includegraphics[width=0.233\textwidth]{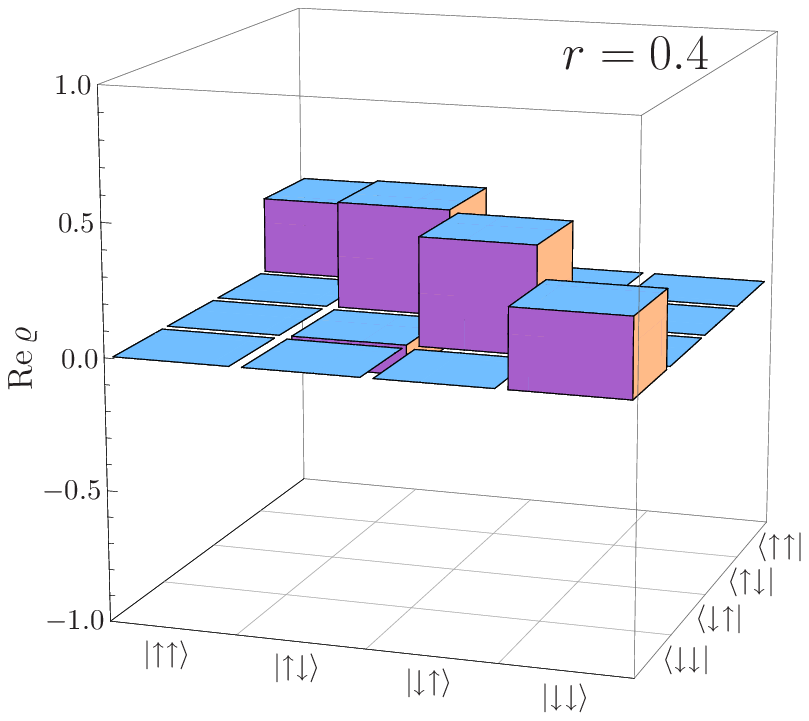}
\end{tabular}
\caption{(Color online) State tomography of $\ket{\Psi^-}=(\ket{\uparrow\downarrow}_\text{XA}-\ket{\downarrow\uparrow}_\text{XA})/\sqrt{2}$ with partially polarized spin channel. The polarization of the channel is defined by $r=(N_\uparrow-N_\downarrow)/(N_\uparrow+N_\downarrow)$ with $N_{{\uparrow}(\downarrow)}$ the number of spins in the  $\ket{{\uparrow}(\downarrow)}$ state in the channel.}
\label{fig:TomoSing}
\end{figure}

\begin{figure}
\begin{tabular}{cc}
\includegraphics[width=0.233\textwidth]{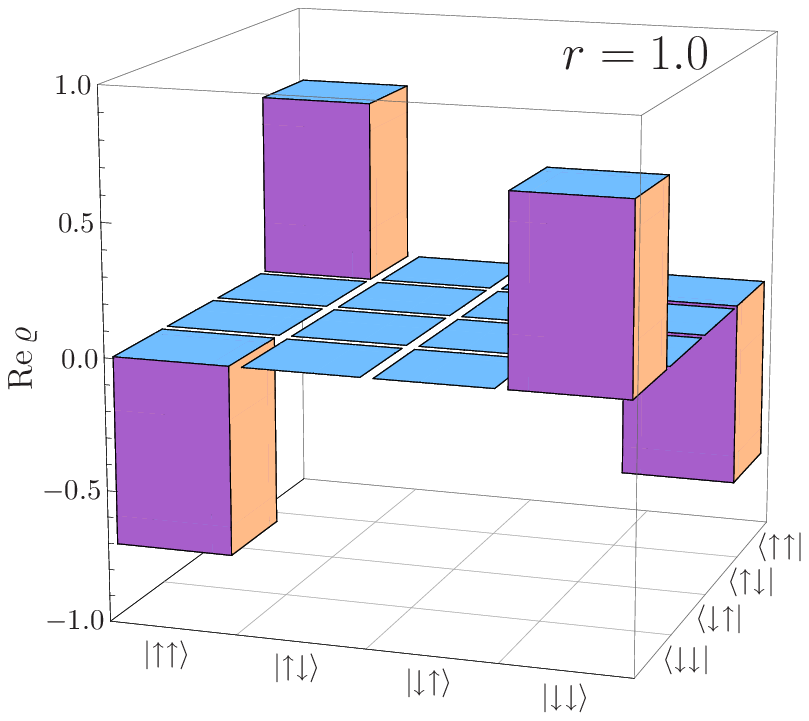}&
\includegraphics[width=0.233\textwidth]{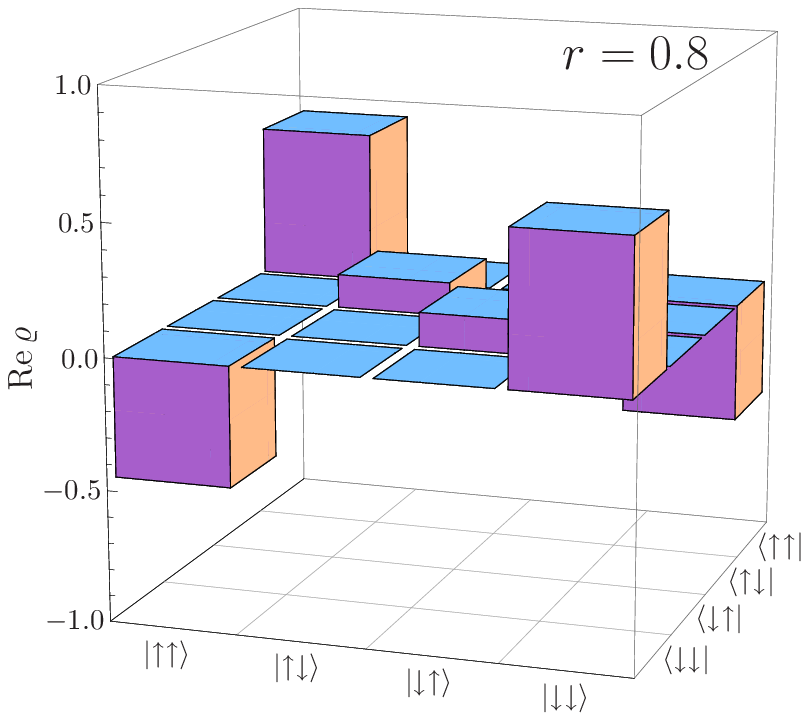}\\
\includegraphics[width=0.233\textwidth]{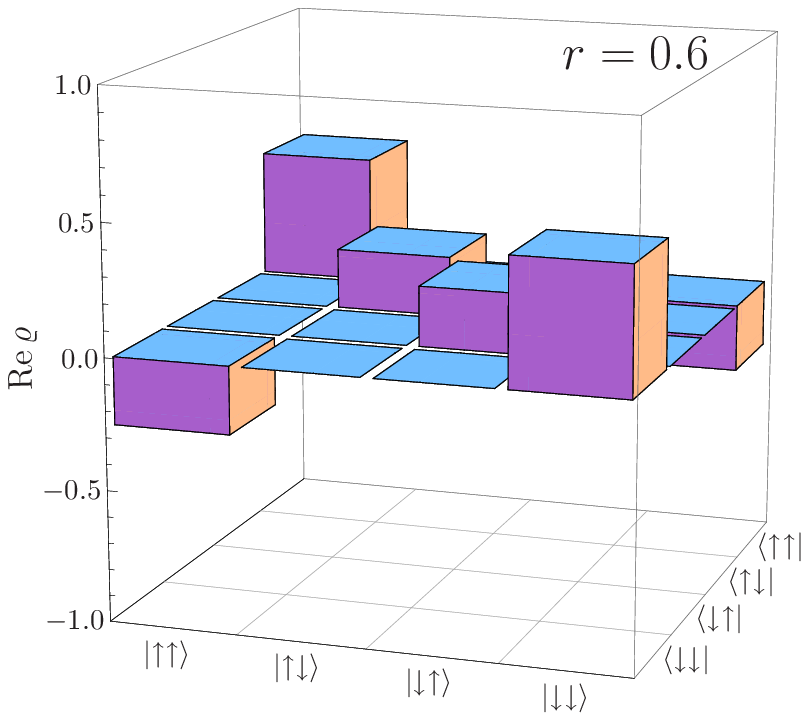}&
\includegraphics[width=0.233\textwidth]{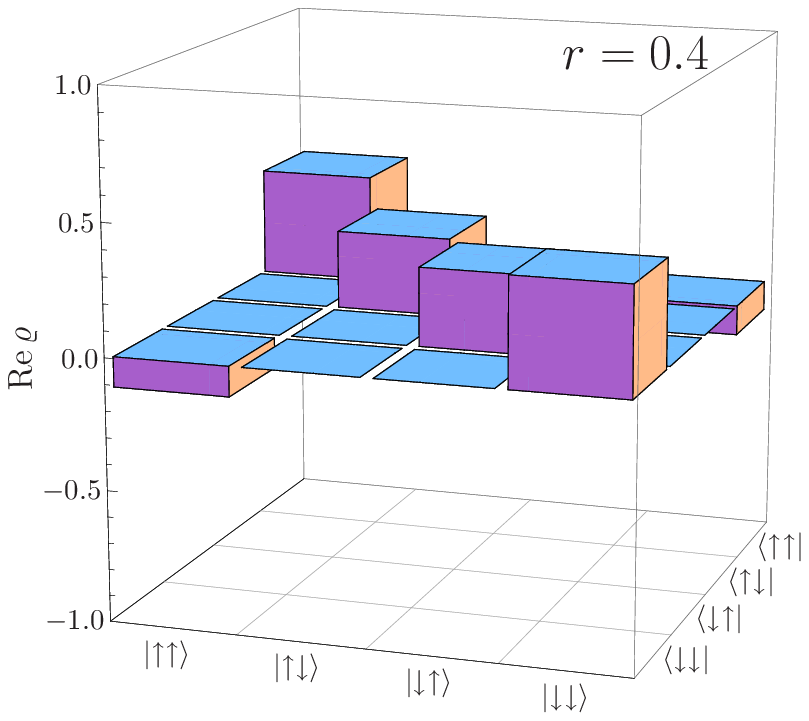}
\end{tabular}
\caption{(Color online) State tomography of $\ket{\Phi^-}=(\ket{\uparrow\uparrow}_\text{XA}-\ket{\downarrow\downarrow}_\text{XA})/\sqrt{2}$ with partially polarized spin channel.}
\label{fig:TomoTrip}
\end{figure}

See Figs.\ \ref{fig:TomoSing} and \ref{fig:TomoTrip}, where the state reconstructions are simulated with the spin-blockade measurement with partially polarized spin channel, Eq.\ (\ref{eqn:NonidealMeas}), in place of the ideal projective measurement $P_\uparrow$ in the recipe presented in Table \ref{tab:SeqTwoQubits}\@.

We have considered the imperfection of the spin-blockade measurement due to the partially polarized spin channel.
The present formalism [Eq.\ (\ref{eqn:NonidealMeas})] would however be applicable to other physical origins of imperfections, including imperfect Pauli blocking.\cite{ref:DecoSemicondDot} The meaning of the parameter $r$ is just to be reinterpreted.

\section{Towards Multi-Qubit Systems}
\label{sec:MultiQubits}
Towards the tomography of multiple qubits (with $M\ge3$), there exists a scheme applicable for an arbitrary number of qubits $M$.
Here is the recipe: we consider four sequences of operations
\begin{equation}
R_{\theta_0\varphi_0}^\text{(X)}
\to P_\uparrow
\,[\to U(\tau)\to R_{\theta\varphi}^\text{(X)}\to P_\uparrow]^{N-1}
\end{equation}
with $(\theta_0,\varphi_0)=(0,0),(\pi,0),(\pi/2,0),(\pi/2,\pi/2)$ and $N\le D^2/4$, 
where $D=2^M$ is the dimension of the $M$-qubit system and $R_{\theta\varphi}^\text{(X)}$ represents a rotation of X defined by
\begin{equation}
R_{\theta\varphi}^\text{(X)\dag}\ket{\uparrow}_\text{X}
=e^{-i\varphi/2}\cos\frac{\theta}{2}\ket{\uparrow}_\text{X}
+e^{i\varphi/2}\sin\frac{\theta}{2}\ket{\downarrow}_\text{X}.
\end{equation}
Let $p_{\theta_0\varphi_0}(N)$ denote the probability of finding X in $\ket{\uparrow}_\text{X}$ successively $N$ times up to the $N$th measurement, which is given by 
\begin{align}
p_{\theta_0\varphi_0}(N)
&=\Tr\{
[P_\uparrow R_{\theta\varphi}^\text{(X)}U(\tau)]^{N-1}
P_\uparrow R_{\theta_0\varphi_0}^\text{(X)}\varrho
\nonumber\displaybreak[0]\\
&\qquad\qquad
{}\times
R_{\theta_0\varphi_0}^\text{(X)\dag}P_\uparrow
[U^\dag(\tau)R_{\theta\varphi}^\text{(X)\dag}P_\uparrow]^{N-1}
\}
\nonumber\displaybreak[0]\\
&=\Tr_\text{AB\ldots}\{[V_{\theta\varphi}(\tau)]^{N-1}\varrho_{\theta_0\varphi_0}[V_{\theta\varphi}^\dag(\tau)]^{N-1}\}
\end{align}
with
\begin{subequations}
\begin{align}
&V_{\theta\varphi}(\tau)
=\bras{\uparrow}{\text{X}}R_{\theta\varphi}^\text{(X)}U(\tau)\ket{\uparrow}_\text{X},
\displaybreak[0]\\
&\varrho_{\theta_0\varphi_0}
=\bras{\uparrow}{\text{X}}R_{\theta_0\varphi_0}^\text{(X)}\varrho R_{\theta_0\varphi_0}^\text{(X)\dag}\ket{\uparrow}_\text{X}.
\end{align}
\end{subequations}
Then, such probabilities are related to the matrix elements of the given density operator $\varrho$ of X+A+B+$\cdots$ through
\begin{equation}
\begin{pmatrix}
p_{\theta_0\varphi_0}(1)\\
p_{\theta_0\varphi_0}(2)\\
\vdots\\
p_{\theta_0\varphi_0}(\frac{D^2}{4})
\end{pmatrix}
=\mathcal{M}
\begin{pmatrix}
\bracket{u_1}{u_1}\bra{v_1}\varrho_{\theta_0\varphi_0}\ket{v_1}\\
\bracket{u_2}{u_1}\bra{v_1}\varrho_{\theta_0\varphi_0}\ket{v_2}\\
\vdots\\
\bracket{u_{\frac{D}{2}}}{u_{\frac{D}{2}}}\bra{v_{\frac{D}{2}}}\varrho_{\theta_0\varphi_0}\ket{v_{\frac{D}{2}}}
\end{pmatrix},
\label{eqn:LinearRel}
\end{equation}
where
\begin{equation}
\mathcal{M}
=\begin{pmatrix}
1&1&\cdots&1\\
\lambda_1\lambda_1^*&
\lambda_1\lambda_2^*&
\cdots&
\lambda_{\frac{D}{2}}\lambda_{\frac{D}{2}}^*\\
(\lambda_1\lambda_1^*)^2&
(\lambda_1\lambda_2^*)^2&
\cdots&
(\lambda_{\frac{D}{2}}\lambda_{\frac{D}{2}}^*)^2\\
\vdots&\vdots&\ddots&\vdots\\
(\lambda_1\lambda_1^*)^{\frac{D^2}{4}-1}&
(\lambda_1\lambda_2^*)^{\frac{D^2}{4}-1}&
\cdots&
(\lambda_{\frac{D}{2}}\lambda_{\frac{D}{2}}^*)^{\frac{D^2}{4}-1}
\end{pmatrix}
\label{eqn:Vandermonde}
\end{equation}
and
\begin{equation}
V_{\theta\varphi}(\tau)\ket{u_n}
=\lambda_n\ket{u_n},\quad
\bra{v_n}V_{\theta\varphi}(\tau)
=\lambda_n\bra{v_n}.
\end{equation}
The non-Hermitian operator $V_{\theta\varphi}(\tau)$ has been assumed to be diagonalizable.\cite{ref:qpfe}

The matrix $\mathcal{M}$ in (\ref{eqn:Vandermonde}) is a Vandermonde matrix of order $D^2/4$, whose properties are well known.\cite{ref:IntegralTable}
In particular, its determinant is given by
\begin{equation}
\det\mathcal{M}
=\prod_{(m,n)>(k,\ell)}(\lambda_m\lambda_n^*-\lambda_k\lambda_\ell^*)
\label{eqn:VandermondeDeterminant}
\end{equation}
and the formula for the inverse $\mathcal{M}^{-1}$ is available, where $(m,n)>(k,\ell)$ means $I_{mn}>I_{k\ell}$ with $I_{mn}=(D/2)(m-1)+n$.
The determinant (\ref{eqn:VandermondeDeterminant}) is the product of all the differences that can be formed by any pairs taken from $\{\lambda_m\lambda_n^*\}$. 
It is therefore clear when it is possible to invert the relation (\ref{eqn:LinearRel}) to reconstruct the density operator $\varrho_{\theta_0\varphi_0}$ of A+B+$\cdots$: the parameters $\tau$ and $(\theta,\varphi)$ should be chosen so as to satisfy the conditions
\begin{equation}
\begin{cases}
\medskip
\lambda_m\lambda_n^*\bracket{u_n}{u_m}\neq0\ (m,n=1,\ldots,D/2),\\
\lambda_m\lambda_n^*\neq\lambda_k\lambda_\ell^*\quad\text{for}\quad(m,n)\neq(k,\ell).
\end{cases}
\label{eqn:ConditionMulti}
\end{equation}
In this way, one gets a list of $\varrho_{\theta_0\varphi_0}$ for the four independent sets of $(\theta_0,\varphi_0)$, which completes the tomography of the state $\varrho$.

This scheme is quite simple and general: one simply repeats $P_\uparrow R_{\theta\varphi}^\text{(X)}$ necessary times, the scheme works for an arbitrary number of qubits $M$, and the conditions for the parameters are clear [Eq.\ (\ref{eqn:ConditionMulti})].
Only four independent sequences are required, irrespective of the number of qubits $M$.

\section{Summary}
In this article, we have discussed the state tomography for a chain of qubits in the setup (Fig.\ \ref{fig:FET}) proposed as a possible physical realization of a quantum information processor.
In this setup, only the state of the edge qubit of the chain is measurable via the spin-blockade measurement.
However, the present analysis explicitly demonstrates that it is still possible to reconstruct the state of the whole chain of qubits.
The idea is to make use of the entangling dynamics of the qubits, which enables one to gain information on the whole chain through the edge qubit.

Such an idea is not restricted to the current setup: there would be various physical systems in which only limited degrees of freedom are accessible in practice and similar strategies are required.
The present idea would find many valuable applications.

In Sec.\ \ref{sec:idea}, we have explicitly constructed a recipe for two qubits.
We have also presented a general scheme based on repeated measurements that can be applied to an arbitrary number of qubits (Sec.\ \ref{sec:MultiQubits}).
There are, however, many other possibilities.
For instance, the latter scheme requires $2^{2(M-1)}$ measurements for $M$ qubits, i.e.\ 4 measurements for 2 qubits, while each sequence in Table\ \ref{tab:SeqTwoQubits} involves only 2 measurements.
As mentioned in Sec.\ \ref{sec:PartialPolarization}, the accuracy of the spin-blockade measurement relies on the polarization of the channel spins.
For a partially polarized spin channel, it would be better to seek a scheme with fewer measurements.
Shorter sequences of operations would be preferable also to minimize other possible errors, originating for instance from imperfect qubit rotations by ESR and decoherence during the processes.

There would exist recipes that involve only one measurement (and therefore with fewer rotations and a shorter execution time) for each sequence, but no general prescription for generating such sequences of operations is known (at least to the present authors).
A possible strategy would be to generate sequences according to a certain rule anyway, to select necessary number of sequences (since too many sequences might be generated for the reconstruction of a density operator with $4^M-1$ independent matrix elements), and to check the invertibility of the relevant matrix relating the matrix elements of the target density operator to the observable data.
It is desirable to clarify how to generate the optimal sequences efficiently, which remains a future subject.

\acknowledgments
This work is partly supported by the bilateral Italian-Japanese
Projects II04C1AF4E on ``Quantum Information, Computation and
Communication'' of the Italian Ministry of Education, University
and Research, and the Joint Italian-Japanese Laboratory on ``Quantum Information and Computation''
of the Italian Ministry for Foreign Affairs, by a Special Coordination Fund for Promoting Science and
Technology from the Ministry of
Education, Culture, Sports, Science and Technology, Japan, and by
the Grants-in-Aid for Scientific Research (A) and (C) from the Japan Society for the Promotion of Science. 
Support by Japan Science and Technology Agency is also gratefully acknowledged.

\end{document}